\documentstyle[prd,aps]{revtex}

\newcommand{\bea}{\begin{eqnarray}}
\newcommand{\eea}{\end{eqnarray}}

\begin{document}

\draft

\title{Cosmological perturbations in a gravity with
       quadratic order curvature couplings}
\author{Hyerim Noh${}^{(a,c)}$ and Jai-chan Hwang${}^{(b,c)}$}
\address{${}^{(a)}$ Korea Astronomy Observatory,
                    San 36-1, Whaam-dong, Yusung-gu, Daejon, Korea \\
         ${}^{(b)}$ Department of Astronomy and Atmospheric Sciences,
                    Kyungpook National University, Taegu, Korea \\
         ${}^{(c)}$ Max-Planck-Institut f\"ur Astrophysik,
                    Karl-Schwarzschild-Str. 1,
                    85740 Garching bei M\"unchen, Germany}
\date{\today}
\maketitle

\begin{abstract}

We present a set of equations describing the evolution of the scalar-type 
cosmological perturbation in a gravity with general quadratic order 
curvature coupling terms.
Equations are presented in a gauge ready form, thus are ready to implement
various temporal gauge conditions depending on the problems.
The Ricci-curvature square term leads to a fourth-order differential 
equation for describing the spacetime fluctuations in a spatially
homogeneous and isotropic cosmological background.

\end{abstract}

\noindent


{\it 1. Introduction:}
In a series of work \cite{Rab-GW,Rab-rot}
we have been studying the evolution of cosmological perturbations 
in a gravity with additional curvature-square correction terms
\bea
   S = \int d^4 x \sqrt{-g} \left[ {1 \over 2} 
       \left( R + A R^2 + B R^{ab} R_{ab} \right) + L_m \right],
   \label{action}
\eea
where $L_m$ is the matter part Lagrangian.
The gravitational field equation is presented in Eq. (2) of \cite{Rab-GW}.
The one-loop order quantum correction generically leads to curvature-square
terms, and due to Gauss-Bonnet theorem the two terms in Eq. (\ref{action}) 
completely account such corrections, \cite{Quantum-correction}. 
In the literature, many works analyzed the roles of $R^2$ term
in the context of cosmological structure formation, \cite{R^2-pert}.
However, we find rarely any work considering the Ricci-curvature square term
in the same context.
When the curvature-square terms come from quantum correction of 
Einstein gravity, if $R^2$ term is important $R^{ab} R_{ab}$ term should be 
important as well because two terms are of the same order.

As a background world model we consider a spatially flat, homogeneous,
and isotropic, Friedmann-Lema\^itre-Robertson-Walker (FLRW), model.
The general perturbations of the FLRW spacetime can be decomposed into
three-types based on the tensorial properties of the perturbed variables: 
these are the scalar-type, (transverse) vector-type, 
and (transverse-tracefree) tensor-type perturbations.
The tensor- and vector-type perturbations correspond to the gravitational 
wave and the rotational perturbation, respectively.
The scalar-type perturbation includes the density condensation and 
the corresponding longitudinal type velocity perturbation, gravitational
potential (curvature) perturbation, etc.
Due to the high symmetry in the background model (homogeneity and isotropy)
the three-types of perturbations decouple from each other in the linear 
level and evolve independently.
Thus, without losing generality we can consider the three perturbations 
independently.
The gravitational wave and the rotational perturbations
were analysed in \cite{Rab-GW} and \cite{Rab-rot}, respectively.
{}For the gravitational wave we derived a fourth-order differential equation,
whereas, the rotational perturbation is simply described by the conservation
of angular momentum.
In this {\it Brief Report}, we will present a complete set of equations 
for the scalar-type perturbation in a gauge-ready form: 
these are in Eqs. (\ref{1}-\ref{7}).
We will show that, in general, the Ricci-curvature square term leads to a
fourth-order differential equation. 
Applications to specific cosmological scenarios are left for future work.

\vskip .5cm
{\it 2. Scalar-type perturbation:}
The cosmological spacetime metric with the most general scalar-type 
perturbation is
\bea
   ds^2 = - a^2 \left( 1 + 2 \alpha \right) d \eta^2
       - a^2 \beta_{,\alpha} d \eta d x^\alpha
       + a^2 \left[ \delta_{\alpha\beta} \left( 1 + 2 \varphi \right) 
       + 2 \gamma_{,\alpha\beta} \right] d x^\alpha d x^\beta,
   \label{metric}
\eea
where $a(\eta)$ is a cosmic scale factor. 
$\alpha ({\bf x}, \eta)$, $\beta ({\bf x},\eta)$, $\varphi ({\bf x}, \eta)$
and $\gamma({\bf x}, \eta)$ are perturbed order metric variables. 
Except for two degrees of the gauge redundancy, these variables represent 
the scalar-type metric perturbations.
It is convenient to introduce the following combinations 
$\chi \equiv a ( \beta + a \dot \gamma)$ and
$\kappa \equiv 3 \left( H \alpha - \dot \varphi \right) - (\Delta / a^2) \chi$;
$\Delta$ is a Laplacian based on $\delta_{\alpha\beta}$, an overdot
denotes the time derivative based on $t$ where
$dt \equiv a d \eta$, and $H \equiv \dot a / a$.
$\alpha$ is the perturbed parts of the comoving lapse function.
$\varphi$, $\chi$, and $\kappa$ 
are the perturbed parts of the three-space curvature, the shear, and 
minus of the perturbed expansion scalar of the normal frame vector field,
respectively; see Sec. 2.1.1 in \cite{PRW}.
We decompose $R ({\bf x}, t) = \bar R (t) + \delta R({\bf x}, t)$;
unless necessary, we ignore the overbar which indicates the
quantity to the background order.
{}For later use we present:
\bea
   & & R^0_0 = 3 \left( \dot H + H^2 \right)
       - \left( \dot \kappa + 2 H \kappa + 3 \dot H \alpha
       + {\Delta \over a^2} \alpha \right), \quad
       R^0_\alpha = - {2 \over 3} {1 \over a}
       \left( \kappa + {\Delta \over a^2} \chi \right)_{,\alpha},
   \nonumber \\
   & & R^\alpha_\beta = \left( \dot H + 3 H^2 \right) \delta^\alpha_\beta
       - {1 \over 3} \left[ \dot \kappa + 6 H \kappa
       + 3 \dot H \alpha + {\Delta \over a^2}
       \left( 3 \varphi + \dot \chi + H \chi \right) \right] 
       \delta^\alpha_\beta
       - {1 \over a^2} \left( \varphi + \alpha
       - \dot \chi - H \chi \right)^{|\alpha}_{\;\;\;\beta},
   \label{Rab}
\eea
where $0 = \eta$ and $|$ indicates a covariant derivative based on 
$\delta_{\alpha\beta}$.

\vskip .5cm
{\it 3. Gauge strategy:}
The gauge transformation properties of the metric variables are presented
in Sec. 2.2 of \cite{PRW}. 
Under the gauge transformation $\tilde x^a = x^a + \xi^a$ with 
$T \equiv a \xi^0$, we have [see Eq. (17) in \cite{PRW}]:
\bea
   \tilde \alpha = \alpha - \dot T, \quad
       \tilde \varphi = \varphi - H T, \quad
       \tilde \chi = \chi - T, \quad
       \tilde \kappa = \kappa + ( 3 \dot H + \Delta / a^2 ) T, \quad
       \delta \tilde R = \delta R - \dot R T.
   \label{GT}
\eea
The variables $\alpha$, $\varphi$, $\chi$, $\kappa$, and $\delta R$
only depend on the temporal gauge transformation, thus are spatially
gauge invariant; for $\delta R$ we used Eq. (\ref{Rab}), 
see also Eq. (\ref{7}).
We have several choices for the temporal gauge fixing:
the synchronous gauge ($\alpha \equiv 0$),
the uniform-curvature gauge ($\varphi \equiv 0$),
the zero-shear gauge ($\chi \equiv 0$), 
the uniform-expansion gauge ($\kappa \equiv 0$), and
the uniform-$R$ gauge ($\delta R \equiv 0$). 
Unless we consider perturbations in the energy-momentum tensor
we do not have the comoving gauge (or the uniform-field gauge for
the scalar field), the uniform-density gauge, etc., 
which are related to imposing conditions on $\delta T^a_b$.
Except for the synchronous gauge condition, each of other temporal gauge 
conditions {\it completely fixes} the temporal gauge mode.
Thus, a variable in such a gauge condition corresponds to a 
unique gauge invariant combination involving the variable concerned
and the variable used in the gauge condition.
We proposed the following notation for the gauge invariant variables
\cite{PRW}:
\bea
   \varphi_\chi \equiv \varphi - H \chi \equiv - H \chi_\varphi, \quad
       \delta R_\chi \equiv \delta R - \dot R \chi 
       \equiv - \dot R \chi_{\delta R}, \quad
       \delta R_\varphi \equiv \delta R - {\dot R \over H} \varphi
       \equiv - {\dot R \over H} \varphi_{\delta R}, \quad
       {\rm etc.}
   \label{GI}
\eea
In this manner we can construct systematically all possible gauge invariant 
combinations. 
Since there exist several (in principle, infinitely many) different gauge
conditions available for each variable, and correspondingly several
gauge invariant combinations, our notation for the
gauge invariant variable is convenient in practice. 

In the Ricci-curvature square (we call it Ricci-square) gravity we have several
additional temporal gauge conditions.
{}From Eqs. (\ref{Rab},\ref{GT}) we have:
\bea
   & & \delta \tilde R^0_0 = \delta R^0_0 - \dot R^0_0 T, \quad
       \delta \tilde R^0_\alpha = \delta R^0_\alpha
       - {2 \over a} \dot H T_{,\alpha}, \quad
       \delta \tilde R^\alpha_\beta = \delta R^\alpha_\beta
       - {1 \over 3} \dot R^\gamma_\gamma \delta^\alpha_\beta T.
\eea
Thus, a temporal gauge condition among $\delta R^0_0 = 0$, 
$\delta R^{0|\gamma}_\gamma = 0$, and
$\delta R^\gamma_\gamma - \delta R^0_0 = 0$
also completely fix the temporal gauge degree of freedom; 
but, these are combinations of our fundamental gauge conditions, 
because these conditions are the same as
$\delta R - 4 [ H \kappa + (\Delta/a^2) \varphi] = 0$, 
$\kappa + (\Delta / a^2) \chi= 0$, and $H \kappa + (\Delta/a^2) \varphi = 0$,
respectively, 

In the following section we will present the perturbed set of equations 
{\it without imposing} the temporal gauge condition.
In this way, the right to impose the temporal gauge condition can be used as an 
{\it advantage} for handling problems depending on the mathematical 
simplification or the physical interpretation we can achieve \cite{Bardeen,PRW}.
Often, different gauge conditions suit for different problems
and keeping the equations ready for imposing various gauge conditions
(thus, in a gauge ready form) is practically convenient.

\vskip .5cm
{\it 4. Equations in the gauge ready form:}
Equations for the background are presented in Eq. (7) of \cite{Rab-GW}:
\bea
   H^2 + 2 \left( 3 A + B \right) \left( 2 H \ddot H - \dot H^2
       + 6 H^2 \dot H \right) = {1 \over 3} \mu, \quad
       \dot \mu = - 3 H \left( \mu + p \right), \quad
       \dot H = - {1 \over 2 {\cal F}} \left( \ddot {\cal F}
       - \dot {\cal F} + \mu + p \right),
   \label{BG-eqs}
\eea
where ${\cal F} \equiv 1 + 2 ( A + B/3 ) R$,
$\mu = - \bar T^0_0$, and $p = {1 \over 3} \bar T^\gamma_\gamma$;
$T_{ab}$ is the energy-momentum tensor of the additional matter part Lagrangian.
{}The third equation follows from the first two equations.

The perturbed set of equations follows from the gravitational field equation 
in Eq. (2) of \cite{Rab-GW}.
The complete set of equations in the gauge ready form is the following
[we introduce $F \equiv 1 + 2 A R$, thus $\delta F = 2 A \delta R$]:

\noindent
Definition of $\kappa$:
\bea
   & & \kappa \equiv - 3 \left( \dot \varphi - H \alpha \right)
       - {\Delta \over a^2} \chi.
   \label{1}
\eea
Energy constraint:
\bea
   & & \delta T^0_0
       = 2 F \left[ {\Delta \over a^2} \varphi
       + \left( H + {\dot F \over 2F} \right) \kappa
       + {3 H \dot F \over 2F} \alpha \right] - 3 H \delta \dot F
       + \left[ 3 \left( \dot H + H^2 \right)
       + {\Delta \over a^2} \right] \delta F
   \nonumber \\
   & & \qquad
       - B \Bigg\{
       2 \left( H \kappa + {\Delta \over a^2} \varphi \right)^{\cdot\cdot} 
       + 6 H \left( H \kappa + {\Delta \over a^2} \varphi \right)^\cdot 
       + 2 \left[ 2 \left( \dot H - 6 H^2 \right) - {\Delta \over a^2} \right]
       \left( H \kappa + {\Delta \over a^2} \varphi \right)
       + 3 H \delta \dot R
       - \left( 3 H^2 + {\Delta \over a^2} \right) \delta R
   \nonumber \\
   & & \qquad \qquad
       + 6 H \dot H \dot \alpha + 6 \left( - H \ddot H + 2 \dot H^2
       - 9 H^2 \dot H \right) \alpha 
       - \left( 6 \dot H + 5 H^2 \right)^\cdot \kappa
       - {8 \over 3} H {\Delta \over a^2} \left( \kappa
       + {\Delta \over a^2} \chi \right)
       \Bigg\}.
   \label{2}
\eea
Momentum constraint:
\bea
   & & T^0_\alpha
       = {2 \over 3} {1 \over a} \nabla_\alpha \Bigg[ \!\! \Bigg[ \left[
       - F \left( \kappa + {\Delta \over a^2} \chi
       + {3 \dot F \over 2F} \alpha \right)
       + {3 \over 2} \left( \delta \dot F - H \delta F \right) \right]
   \nonumber \\
   & & \qquad
       - B \Bigg\{ - \left( \kappa + {\Delta \over a^2} \chi 
       \right)^{\cdot\cdot} 
       - 3 H \left( \kappa + {\Delta \over a^2} \chi \right)^\cdot
       + \left( 12 H^2 + {\Delta \over a^2} \right) 
       \left( \kappa + {\Delta \over a^2} \chi \right)
       - {3 \over 2} \left( \delta \dot R - H \delta R \right)
   \nonumber \\
   & & \qquad \qquad
       - 3 \dot H \dot \alpha 
       + {3 \over 2} \left( 2 \dot H + 9 H^2 \right)^\cdot \alpha 
       + 3 {\Delta \over a^2} \left[ 2 H \varphi 
       + \left( \dot H - 2 H^2 \right) \chi \right] \Bigg\}
       \Bigg] \! \! \Bigg].
   \label{3}
\eea
Tracefree propagation:
\bea
   & & \delta T^\alpha_\beta - {1 \over 3} \delta T^\gamma_\gamma
       \delta^\alpha_\beta
       = {1 \over a^2} \left( \nabla^\alpha \nabla_\beta
       - {1 \over 3} \delta^\alpha_\beta \Delta \right)
       \Bigg[ \! \! \Bigg[ F \left[ \dot \chi 
       + \left( H + {\dot F \over F} \right) \chi
       - \varphi - \alpha - {\delta F \over F} \right]
   \nonumber \\
   & & \qquad 
       - B \Bigg\{ 
       \left( \dot \chi + H \chi - \varphi - \alpha \right)^{\cdot\cdot}
       - H \left( \dot \chi + H \chi - \varphi - \alpha \right)^\cdot
       - \left[ 8 \left( \dot H + H^2 \right) + {\Delta \over a^2} \right]
       \left( \dot \chi + H \chi - \varphi - \alpha \right)
   \nonumber \\
   & & \qquad \qquad
       + \delta R - 4 \dot H \varphi - 6 \left( \dot H + H^2 \right)^\cdot 
       \chi
       + {8 \over 3} H \left( \kappa + {\Delta \over a^2} \chi \right) 
       \Bigg\} \Bigg] \! \! \Bigg].
   \label{4}
\eea
Raychaudhuri equation:
\bea
   & & \delta T^\gamma_\gamma - \delta T^0_0
       = 2 F \left\{ \dot \kappa
       + \left( 2 H + {\dot F \over 2 F} \right) \kappa
       + {3 \dot F \over 2F} \dot \alpha
       + \left[ 3 \dot H + {3 \over 2F} \left( 2 \ddot F + H \dot F \right)
       + {\Delta \over a^2} \right] \alpha \right\}
   \nonumber \\
   & & \qquad \qquad
       - 3 \delta \ddot F - 3 H \delta \dot F
       + \left( 6 H^2 + {\Delta \over a^2} \right) \delta F
   \nonumber \\
   & & \qquad
       - B \Bigg\{
       - 4 \left( H \kappa + {\Delta \over a^2} \varphi \right)^{\cdot\cdot}
       - 12 H \left( H \kappa + {\Delta \over a^2} \varphi \right)^{\cdot}
       - 4 \left[ 2 \left( \dot H - 6 H^2 \right) - {\Delta \over a^2} \right]
       \left( H \kappa + {\Delta \over a^2} \varphi \right)
       + 2 \delta \ddot R + 6 H^2 \delta R
   \nonumber \\
   & & \qquad \qquad
       - 28 H \dot H \kappa - 6 \left( 2 \dot H + 5 H^2 \right)^\cdot
       \dot \alpha
       - 12 \left[ \left( 2 \dot H + 5 H^2 \right)^{\cdot\cdot}
       + 3 H^2 \dot H \right] \alpha
       + {16 \over 3} H {\Delta \over a^2} \left( \kappa
       + {\Delta \over a^2} \chi \right)
       \Bigg\}.
   \label{5}
\eea
Trace equation:
\bea
   & & \delta T
       = - \delta R - 2 \left( 3 A + B \right)
       \left[ \delta \ddot R + 3 H \delta \dot R
       - {\Delta \over a^2} \delta R
       - \dot R \left( \kappa + \dot \alpha \right)
       - \left( 2 \ddot R + 3 H \dot R \right) \alpha \right],
   \label{6}
\eea
and we have
\bea
   \delta R = - 2 \left[ \dot \kappa + 4 H \kappa + 3 \dot H \alpha
       + {\Delta \over a^2} \left( 2 \varphi + \alpha \right) \right]. 
   \label{7}
\eea
{}We decomposed the equation for 
$\delta T^\alpha_\beta$ into tracefree and trace parts.
Equations (\ref{1}-\ref{7}) form a redundantly complete set 
describing the scalar-type perturbation in a gauge ready form; 
for example, the combination of Eqs. (\ref{2},\ref{5}) leads to Eq. (\ref{6}).
In the limit of $R^2$ gravity (thus $B = 0$)
Eqs. (\ref{1}-\ref{6}) correspond to Eqs. (54-58,60) in \cite{GGT-HN} 
which also correspond to Eqs. (22-26) in \cite{PRW};
the names of the equations are based on our convention in \cite{PRW}.
When we analyse we have the right to impose one 
temporal gauge condition out of several choices displayed in Sec. {\it 3}.

\vskip .5cm
{\it 5. $R^2$ gravity:}
We set $B = 0$. 
Without additional matter terms ($T_{ab} = 0$), the analysis can be made 
most conveniently in terms of $\varphi$ in the uniform-$R$ gauge where
we take $\delta R \equiv 0$ (thus $\delta F = 0$) as the gauge condition;
equivalently, we let $\delta R = 0$ and change the other
perturbation variables into the gauge invariant forms, for example,
$\varphi \rightarrow \varphi_{\delta R}$, etc.
Using Eq. (\ref{1}-\ref{3}) we express $\kappa$ and $\alpha$ 
in terms of $\varphi$.
{}From Eq. (\ref{6}) we can derive
\bea
   {( H F / \dot F + 1 / 2 )^2 \over a^3 F}
       \left[ {a^3 F \over ( H F / \dot F + 1 / 2 )^2}
       \dot \varphi_{\delta R} \right]^\cdot
       - {\Delta \over a^2} \varphi_{\delta R} = 0.
   \label{R^2-eq}
\eea
This equation was derived in a more general context 
in \cite{GGT-Unified,GGT-HN};
it is valid for a general action with $f(R)$ replacing
$R + A R^2$ with $F \equiv \partial f/(\partial R)$.
Perturbation analyses in the $R^2$ gravity were also pursued 
in \cite{R^2-pert} in different gauge conditions.
Using $\varphi_{\delta R} = - (H / \dot R) \delta R_\varphi$
in Eq. (\ref{GI}) we can derive the equation for $\delta R_\varphi$.
In the large scale limit (thus, ignoring the Laplacian term) we have 
an integral form {\it solution valid for the generally time varying background 
dynamics} 
\bea
   \varphi_{\delta R} ({\bf x}, t) 
      = - {H \over \dot R} \delta R_\varphi ({\bf x}, t)
      = C ({\bf x}) + {2 \over 3} D ({\bf x}) \int_0^t
      { ( H F / \dot F + 1 / 2 )^2 \over a^3 F } dt,
   \label{conservation}
\eea
where $C({\bf x})$ and $D({\bf x})$ are integration constants indicating
the growing and decaying solutions, respectively; the decaying solution is
higher order in the large scale expansion, see Eq. (109) in \cite{GGT-HN}.
Thus, the non-transient solution of $\varphi_{\delta R}$
is conserved in the large scale limit, and the generalized nature of the
gravity does not affect this result!

\vskip .5cm
{\it 6. Discussions:}
In general we expect the curvature square terms lead to fourth-order
gravity theories.
It happens that, as we just saw, the $R^2$ or $f(R)$ gravity leads to 
a second-order system.
This is probably due to the conformal symmetry of the $f(R)$ gravity to the
Einstein gravity with a minimally coupled scalar field, \cite{CT}.
The Ricci-square term does not have such a conformal symmetry, and
we expect to have a fourth-order differential equation for the scalar-type
perturbation.
In general, we do not know which gauge condition will be the most
suitable for the problem {\it a priori}.
As a trial attempt we can impose the same gauge condition used in $R^2$ 
gravity which is the uniform-$R$ gauge; thus, we set $\delta R \equiv 0$ 
and $\delta F = 0$.
One possible way to derive a fourth-order equation is the following.
Using Eqs. (\ref{1},\ref{6},\ref{7}) we can express $\alpha$, $\dot \alpha$, 
and $\kappa + (\Delta/a^2) \chi$ in terms of $\kappa$ and $\varphi$.
{}From a time derivative of Eq. (\ref{7}), and from Eq. (\ref{2}) 
we can derive a coupled set of two second-order differential 
equations for $\kappa$ and $H \kappa + (\Delta/a^2) \varphi$. 
In such a special form, however, it happens that the equations alone cannot 
be reduced to Eq. (\ref{R^2-eq}) in a pure $R^2$ gravity limit.

Our {\it main result} is Eqs. (\ref{1}-\ref{7}) which describe the evolution of
a perturbed flat FLRW background in a gauge ready form;
the evolution of background is governed by Eq. (\ref{BG-eqs}).
The set of equations for the Ricci-square term is a new result, 
and after imposing a suitable gauge condition it generally leads to
a fourth-order differential equation without remaining gauge mode.
We mentioned one possible set of a fourth-order differential equation in 
a previous paragraph as a demonstration. 
However, the mentioned set may not be the best form possible
in the available pool of the variables and the gauge conditions in Sec. {\it 3}.
Equations (\ref{1}-\ref{7}) are arranged in a general gauge-ready form, 
thus allow an easy implementation of one's favored choice for the variables 
and the temporal gauge.

The fourth-order terms in Eq. (\ref{action}) usually appear as the 
one-loop order quantum corrections. 
In this regard such terms can be regarded as a transient 
(because higher-order correction terms will appear soon in higher order 
quantum corrections) form of medium-energy effective action 
\cite{Quantum-correction}.
Thus, we regard the $A$ and $B$ terms in Eq. (\ref{action})
as corrections to Einstein action, and in such a case,
we should consider the third and fourth order derivatives as
corrections to the second order equation based on Einstein gravity.
In the case we should accept Eq. (\ref{action}) as the
fundamental theory we are not certain whether the possible four solutions of 
the expected fourth-order differential equation in a general situation 
(and also the fourth-order gravitational wave equation derived in 
\cite{Rab-GW}) should be interpreted as representing physical degrees of 
freedom; for example, we should consider the Cauchy problem and 
the stability of the full theory.
We will not address this important question in this {\it Brief Report}, 
\cite{Fourth-order}.

We briefly describe how we can connect specific solutions of our formulation
with the present day observable quantities.
Since we have some conceptual difficulties mentioned in the previous
paragraph for fourth-order gravity, let us consider
the $R^2$ gravity as an example.
Let us consider a scenario where the higher-order gravity dominates
the early evolution stage of our observable patch of the universe, and
at some point Einstein gravity takes over the dominance till the present era.
If we further assume that the higher-order gravity era provides an
accelerated expansion (inflation) stage, the observationally relevant 
scales may transit from subhorizon to superhorizon scales during the stage.
If conditions (e.g., allowing an analytic form mode-function solution of 
the perturbed equation) are met, we can derive the generated quantum 
fluctuations (usually, in general scales) based on vacuum expectation value,
and as the scale becomes superhorizon it can be interpreted as the classical
fluctuations based on the spatial average.
As long as the relevant scales remain in the superhorizon stage during the
transit epoch of the gravities (from higher-order to Einstein) there exists a 
{\it conserved quantity}.
$\varphi$ in the comoving gauge is the conserved quantity in the superhorizon
scale independently of the changing gravity (as long as the gravity
belongs to a certain class), changing field potential (for the scalar field), 
and changing equation of state (for the hydrodynamic situation), 
see \cite{GGT-Unified}.
In our case, $\varphi_{\delta R}$ in Eq. (\ref{conservation}) is the conserved
quantity; the growing solution is conserved as $C({\bf x})$.
Since we are considering the linear perturbation all quantites are related to 
each other by linear combinations; 
for the specific relations to the quantities with observational consequences
(e.g., temperature anisotropies in the cosmic microwave background radiation, 
and the Newtonian hydrodynamic variables) see Eq. (23) in \cite{GGT-Unified}.
In summary, the curvature fluctuation $\varphi$ in a certain gauge is 
conserved and connects perturbations in both eras.
In later Einstein era with hydrodynamic contents, the density fluctuations 
will fall-in to the potential well of, thus co-grow with, 
the curvature fluctuation.
Concrete analyses of the quantum generation and classical evolution processes
with the generated observational specta in various scenarios based on
a class of generalized gravity theories were carried out in \cite{GGT-Quantum}.
Applications of our formulation to specific cosmological models 
are left for the future work.

\vskip .5cm
JH was supported by the KOSEF, Grant No. 95-0702-04-01-3  
and through the SRC program of SNU-CTP.
HN was supported by the DFG fellowship (Germany) and the KOSEF (Korea).


\end{document}